# Climate Change and Potential Demise of the Indian Deserts

## P. V. Rajesh[1] and B. N. Goswami[2]*

[1]Indian Institute of Tropical Meteorology, Pune 411008, India

[2]Department of Physics, Cotton University, Guwahati 781001, India

*Corresponding author, email bhupengoswami100@gmail.com

## Abstract

In contrast to the "wet gets wetter and dry gets drier" paradigm, here, using observations and climate model simulations, we show that the mean rainfall over the semi-arid northwest parts of India and Pakistan has increased by 10–50% during 1901–2015 and is expected to increase by 50–200% under moderate greenhouse gas (GHG) scenarios (e.g., SSP2-4.5). The GHG forcing primarily drives the westward expansion of the Indian summer monsoon (ISM) rainfall and is a result of a westward expansion of the inter-tropical convergence zone (ITCZ), facilitated by a westward expansion of the Indian Ocean warm pool. While an adaptation strategy to increased hydrological disasters is a must, harvesting the increased rainfall would lead to a significant increase in food productivity, bringing transformative changes in the socio-economic condition of people in the region.

## 1 Introduction

An iconic feature of the South Asian monsoon is its east-west asymmetry, with the Great Indian "Thar" Desert at the center of the semi-arid northwest India compared to the highly moist northeast India (Fig. 1a). The interaction of low-level southwesterly monsoon winds and the northern Indian Ocean (IO) to the south of it, resulting in cooling of the western part of the Arabian Sea (AS), is critical in maintaining the aridity of northwest India (Gadgil, 2003; Goswami & Goswami, 2017; Goswami & Chakravorty, 2017; Webster et al., 1998). With an annual rainfall amount of 40–50 cm compared to more than 250 cm in the eastern part of India, this region is particularly vulnerable to year-to-year variations in the mean monsoon rainfall and experiences an increasing trend of daily extreme rainfall events (Goswami, Madhusoodanan, et al., 2006; Rajeevan et al., 2008). Evidence from paleoclimatic and archaeological data shows that the Indus Valley Civilization (5300–3300 yr. BP) and civilizations during the early (3400–3050 yr. BP) and later (3050–2450 yr. BP) Vedic periods flourished around the Indus River (70° ± 2°E), extending only as far as central India, indicates a significant spatial shift of the monsoon core (Kathayat et al., 2017) and sustained precipitation zone towards the west. Similar archaeological evidence of the Indus Valley Civilization between the border regions of India and Pakistan has also shown pre-existing and more extensive signs of civilization over the Thar Desert, associated with the revival of the river Saraswati (Kineman & Anand, 2015), suggesting that that northwest India was not always this arid in the past.

Interestingly, many modeling studies indicate that global warming (GW) is expected to cause precipitation to decrease in most semi-arid regions, and vegetation-albedo feedback is likely to lead to the expansion of the global deserts (Koutroulis, 2019; Liu & Xue, 2020; Wetherald & Manabe, 2002; Zeng & Yoon, 2009). However, the response of the South Asian monsoon semi-arid zone to GW could be quite different if it forces a westward expansion of the South Asian monsoon like the westward expansion of "land monsoon" expected under GW (Lee & Wang, 2014). Under such circumstances, expected and enduring ramifications of climate change over the semi-arid northwest region of India, therefore, have positive transformative repercussions for the region's socio-economic conditions and implications for the expansion of deserts in the warming world (Spinoni et al., 2013; Zdruli et al., 2016).

The future changes in the length of the rainy season (LRS) and the changes in the zonal extent of the "core monsoon zone" (Rajeevan et al., 2010) are still unexplored, even though we are now beginning to understand how climate change is expected to impact the characteristics of the ISM rainfall (Turner & Annamalai, 2012; Wang et al., 2021). Between 1901 and 2005, the intensity of daily extreme rainfall controlled by local small-scale instability increased at an overwhelming rate of 29.5% $K^{-1}$ over the country (Rajesh et al., 2021) and 12.5% $K^{-1}$ over north-east India (Zahan et al., 2021), owing to an increase in mean atmospheric moisture content at nearly the Clausius-Clapeyron rate as a result of rapid GW. However, the seasonal mean rainfall, governed by large-scale dynamics is found only to increase at 3.95% $K^{-1}$ over the country (Rajesh et al., 2021) between 1850 and 2005, while north-east India it decreases at -3.2% $K^{-1}$. Over the region, GW is also associated with an anomalously strong increasing trend of sea surface temperature (SST) over the tropical IO, with the most significant increase over the western IO and AS (Roxy et al., 2014, 2015), when compared to the other oceanic basins.

The spatially non-uniform warming of the IO has led to an expansion of the warm pool towards the west( Roxy et al., 2017; Weller et al., 2016), resulting in more intense cyclogenesis in the eastern part of the AS (Deshpande et al., 2021) over recent years. While the cause of the expansion of the Indo-Pacific warm pool is still being debated, human-induced greenhouse gas forcing appears to be the major driver (Weller et al., 2016) for such a warming. Climatologically, the active ITCZ supported by the warm pool is restricted to about 75°E, which limits moisture supply to mesoscale synoptic activity in northwestern India. Due to this constraint, climatologically, the rainy season in NW India is relatively "short," lasting about 70 days, compared to the "long" rainy season in NE India, which lasts about 150 days (Misra et al., 2018). The expansion of the IO warm pool to the west associated with GW has allowed the active ITCZ to expand to at least 65°E and makes more moisture and instability available to northwest India, thereby could potentially increase the LRS over northwest India in recent decades. Therefore, if the increase in GW is expected to increase the LRS in northwest India, it will result in a further westward expansion of the ISM season and any significant change in the rate at which the seasonal mean rainfall over the semiarid northwest India may have a transformative impact in the region. Here, we provide evidence that the northwest India is getting significantly wetter in recent years.

Using observations and the ensemble of the latest historical CMIP6 (Coupled Model Inter-comparison Project-Phase 6) simulations for the historical period 1901–2015, here we show an unambiguous westward expansion of the core monsoon zone with up to a 25–50% increase in rainfall over the semi-arid west and a slighter decrease over the wet northeast India during the period (1901–2015). We also show that the overall spatial pattern of increased rainfall to the west remains the same under increasing greenhouse gas forcing, while the magnitude intensifies 2–3 times larger when compared to the historical changes.

## 2 Data and Methods

### 2.1 Observed data

Rainfall: Monthly rainfall data from the GPCP from 1985 to 2015 is used to estimate the current rainfall climatology and model bias over the South Asian monsoon domain, including oceanic regions. Fixed stations-based daily mean rainfall from the IMD 1°×1° gridded data (Rajeevan et al., 2010) have been used to estimate past and present historical climatology as well as changes in mean rainfall over the Indian subcontinent.

SST: The HadISST1.1 SST dataset with a grid resolution of 1°×1° is used to derive the long-term June-Sept (JJAS) and Dec-Feb (DJF) SST climatology as well as to estimate the model bias in SST. The dataset is considered accurate enough for studies of the large-scale signals associated with short-term climate variability (Chelton & Risien, 2016).

MSE: ERA5 provided wind, temperature, and geopotential fields are used to derive the cloud-base (950 hPa) MSE. These fields, together with 850 hPa winds, were used to calculate the trends in the present period (1979 through 2019). As the trends of long-term low-level winds have significant changes among different historical datasets, we restrict ourselves to a shorter (40-year) ERA5 monthly reanalysis, which is one of the best satellite assimilated datasets.

LRS: The LRS of ISM was computed by obtaining the difference between local rainfall onset and cessation dates for the historical period (1901–2019) and evaluating it using IMD data based on the methods of Misra(Misra et al., 2018), where any bogus local onset and withdrawal days for those years are discarded while estimating the trends. The above-mentioned dataset is also used to estimate the frequency distributions of daily rainfall and their epochal variations over central-eastern India (CI, 80°E–95°E, 15°E–25°N) and northwest India (NWI, 65°E–80°E, 15°N–35°N). The LPS data is taken from the Monsoon Low-Pressure System Global Track dataset (Vishnu et al. (2020), ERA reanalysis). The LPS rainfall amount and LPS duration are derived from IMD 1°×1° data using the corresponding LPS track data. The rank-based non-parametric Mann-Kendall's test is applied to the hydrologic and atmospheric variables to estimate the trend significance and the slope of the trends.

2.2 Model Data

We used the rainfall and SST monthly historical simulations as well as climate projections from a common suite of 34 CMIP6 (Table S1) global climate models available for fossil-fuel-rich development Shared Socio-economic Pathway (SSP2–4.5 and SSP5–8.5) scenarios and 30 CMIP5 (Table S1) models under historical simulations and Representative Concentration Pathways (RCP8.5 and RCP4.5) scenarios. Model outputs of monthly mean precipitation and SST fields are obtained for the first realization (r1i1p1f1) from CMIP6 models as well as (r1i1p1) from CMIP5 models. The SNR as defined by the ratio of the magnitude of the change in predicted trend (MME mean change) to inter-model standard deviation is used to measure the inter-model dispersion, and the inter-model agreement is depicted by overlaying dots over the region where the SNR is greater than 1.

Estimation of climatological changes

In the present study, the changes in precipitation (also, percentage changes) and changes in SST during the historical period as well as the projections are estimated based on the reference late pre-industrial baseline climatology period (1901–1930) ($PR_{past}$). Accordingly, the climatological changes during the historical period are therefore calculated by subtracting the past baseline climatology ($PR_{past}$ 1901–1930) from the present historical climatology ($PR_{pres}$ 1985–2014). Similarly, the projected changes are calculated as the difference between $PR_{hist}$ and the climatology of the last 30 years in the projected period ($PR_{proj}$, 2071–1999). The differences are computed for both the boreal summer months (JJAS) as well as the boreal winter months (DJF).

Estimation of the best CMIP6 models

The best 10 CMIP6 models for simulating the ISM rainfall were chosen based on their ability to reproduce a reasonable amount of seasonal mean, spatially averaged seasonally accumulated rainfall greater than 600 mm (observed long term mean is 890mm), and a similar observed spatial pattern. The agreement in the spatial pattern is determined by calculating the correlation coefficients between simulated and observed climatological mean rainfall over the ISM region, which was chosen greater than 0.75.

Correction of ISM using EC

We use the EC (Rajesh & Goswami, 2022) to constrain the spatially averaged mean ISM rainfall estimates using the SSP5–8.5 scenario. Our main objective is to determine whether the EC produced from the same set of 28 CMIP6 models under the SSP5–8.5 scenario would be similar in capturing the rain belt's westward expansion. However, it should be noted that the EC is primarily linked to the monsoon domain over the Indian subcontinent, which is the active monsoon core zone over land, and is bounded to the west by the semi-arid dry regions. The projected changes in rainfall away from monsoon core and towards the far western Arabian Desert regions should therefore be interpreted with some caution. Based on the EC, corrections have been applied for individual models before computing the revised MME changes through the relationships obtained between simulated projected ISM changes (percentage) and simulated historical ISM regressions over the North Atlantic SST (refer Rajesh & Goswami, 2022), for more details about the EC).

## 3. Data and materials availability:

The observed global monthly rainfall data is obtained from GPCP https://psl.noaa.gov/data/gridded/data.GPCP.html and ISM rainfall from IMD 1°×1° daily rainfall data available through http://imdpune.gov.in/Clim_Pred_LRF_New/Grided_Data_Download.html. HadISST1 monthly SST data is available at https://www.metoffice.gov.uk/hadobs/hadisst/data/download.html. All the model historical and projection datasets are downloaded through the web interfaces https://esgf-node.llnl.gov/search/cmip5 , for CMIP5 models and https://esgf-node.llnl.gov/search/cmip6 for CMIP6 models. Climate Data Operators (CDO) open-source software is used for all model dataset computations, while NCL is used for making visualisations.

Code availability

Basic data and codes for generating the figures are available at https://zenodo.org/record/6884537#.YtqzLIRByUl

## 4 Results

### 4.1 Westward Expansion of the South Asian Monsoon Rainfall

The rainfall data between 1979 and 2015 reveal the critical feature of the observed June-September (JJAS) ISM rainfall climatology, which is a gradual reduction in rainfall from the east to the west that results in a semi-arid northwest India (Fig. 1a). Although there is still a significant dry bias across the core monsoon region of central India, the spatial distribution of the simulated ensemble mean JJAS rainfall climatology of 34 CMIP6 models is similar to the observed climatology (Fig. S1a, pattern correlation r =0.82) (Choudhury et al., 2021). The changes in monsoon rainfall during the simulated historical period are calculated by subtracting the climatology averaged spatial pattern of JJAS rainfall MME of the late pre-industrial 30 year period (PR$_{past}$) (1901–1930) from the present

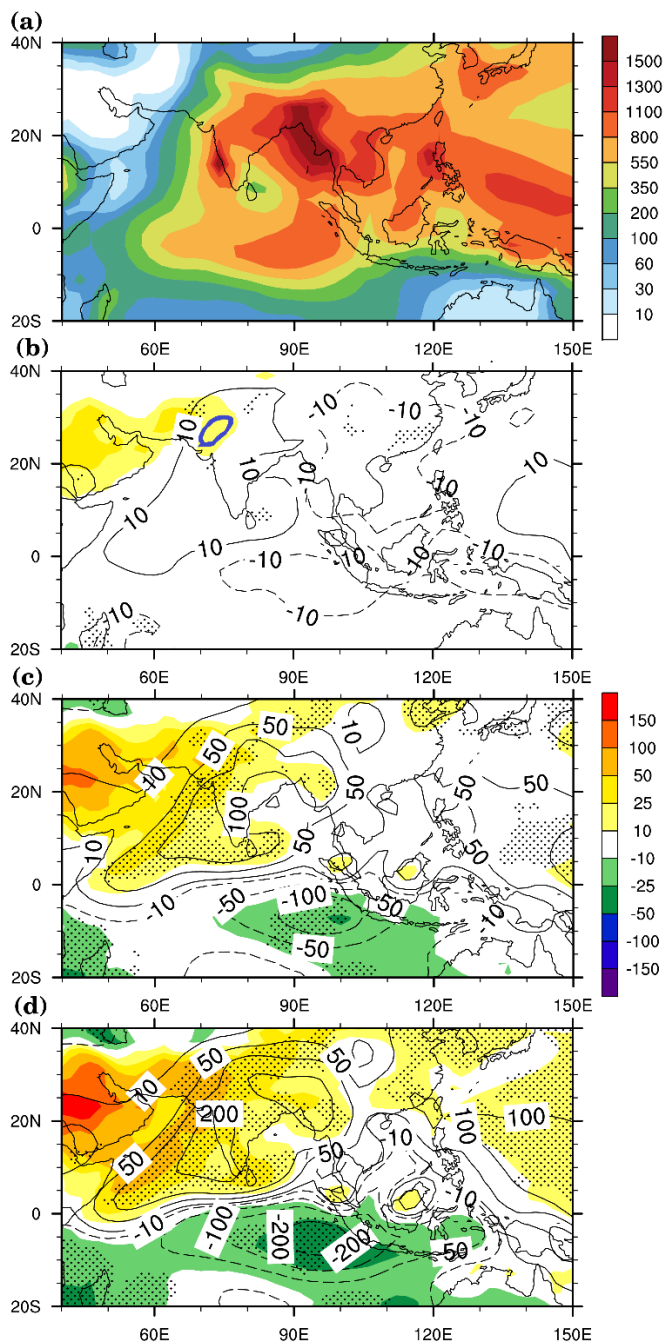

**Fig 1. Present and future changes of the South Asian Monsoon Rainfall** (a) Present-day observed JJAS accumulated rainfall climatology (mm) from the Global Precipitation Climatology Project (GPCP) (shading) (1979–2015). The changes in the CMIP6 simulated MME rainfall amount (black contours), by subtracting the late pre-industrial climatology (PR$_{past}$) (1901–1930) respectively from (b) the present climatology (PR$_{pres}$) (1985–2015) of historical ensemble; (c) from projected climatology at the end of the century (PR$_{proj}$) (2070–2099) for the SSP2–4.5 ensemble and (d) from PR$_{proj}$ at the end of the century for SSP5-8.5 ensemble. The thick blue schematic curve in (b) symbolises the geographical location of the Thar desert. The colour shading in (b, c, and d) represents the percentage change (%) of rainfall amount with reference to the past climatology (PR$_{hist}$), and the contours represent the changes in seasonal accumulated rainfall amount (mm). The dotted area denotes the regions where the magnitude of the projected trend (MME mean change) to signal-to-noise ratio (SNR) (inter-model standard deviation) is greater than one. The list of the 34 CMIP6 models is provided in Table-S1.

30 years (PR$_{pres}$) (1985–2014) of CMIP6 model's historical simulations (Fig. 1b). Similarly, the changes in projected JJAS mean rainfall is estimated by subtracting the climatology of PR$_{past}$ from the last 30 years in the future projections (PR$_{proj}$) (2071–1999) under shared socioeconomic paths of both SSP2–4.5 and SSP5–8.5.

From the evaluation of the rainfall changes during the historical period, the multi-model ensemble mean (MME) climatology exhibits a striking east-west dipole pattern, with mean rainfall increasing by up to 25–50% north of 5°N and west of 95°E while decreasing by a similar amount east of 95°E up to roughly 130°E. The relatively small percentage change in mean rainfall is caused by the significant large mean rainfall toward the east, which suggests that the ISM wet zone is expanding westward. On the other hand, the oceanic rain belt between the equator and 10°S show weakening, with a decrease in mean rainfall from the eastern IO extending up to the maritime continent. These changes in the land and ocean precipitation found by us using CMIP6 models in the context of the South Asian monsoon are similar to those found by Lee and Wang(Lee & Wang, 2014) in the context of global monsoon by CMIP5 models.

As greenhouse gas forcing increases (e.g. SSP2–4.5 and SSP5–8.5), the magnitude of decreasing rainfall zone over the east weakens, while rainfall increases to the northwest of the equator and decreases to the southeast of the equator are both enhanced (Fig. 1c, d). This result is a 50-100% increase in the mean summer rainfall over the semi-arid northwest of India.

4.2 Westward Expansion of the Indian Ocean Warm Pool

The climatological June-September (JJAS) SST over the IO is also generally associated with an east-west asymmetry, with the IO warm pool east of 65°E and north of 10°S but colder waters to the west and south of it (Fig. 2a). With the increasing monsoon rainfall projections over north India (Fig. 1b, c&d), there is an equivalent increase in the SST towards the western IO (Fig. 2c&d). This implies that the westward expansion due to the uneven warming of the IO warm pool( Roxy et al., 2014, 2015)  is intrinsically linked with the westward expansion of the South Asian monsoon rainy zone. According to the observations, there is an uneven increase in the mean climatology of JJAS SST between SST$_{pres}$ (1985–2014) and SST$_{past}$ (1901–1930), rising by 1°C to the west and 0.5°C to the east (Fig. 2b). This east-west asymmetry and the as well as the observed JJAS SST changes (Fig. 2b) are both reproduced by the ensemble mean of 34 CMIP6 historical models, but the magnitude is a little weaker with an increase of 0.3°C in the east and a decrease of 0.6°C in the west (Fig. 2c).

The westward expansion of the warm water blob over the equatorial IO was adequately captured by the MME of the CMIP6 historical models (Fig. S2a), which is consistent with the observed SST climatology (Fig. 2a). However, for the projections, the increase in SST over the western IO increases significantly from 0.6°C in the historical period to 2.5°C in SSP2–4.5 (Fig. 2d) and 3.0°C in SSP5–8.5 (Fig. S2d) while the strength of the east-west asymmetry in SST change increases consistently. The slow rate of warming of the eastern IO and western Pacific warm pool, compared to the faster warming of the western IO, suggests that the IO warm pool will continue to expand westward. The east–west asymmetry of the north IO climatological mean SST during the Boreal summer is intimately linked with the strength of the South Asian summer monsoon. A more vigorous monsoon is associated with a stronger "westerly jet" and stronger south-westerlies over the AS that cools the western north IO SST through Ekman upwelling near the coast and enhances turbulent mixing in the north AS. While external anthropogenic forcing appears to be the dominant driver of the westward expansion of the warm pool (Weller et al., 2016), local ocean-atmosphere interactions could also contribute to it. With GW, the monsoon circulation is likely to weaken due to the increased stability of the atmosphere brought on by upper atmospheric heating and a decrease in the meridional gradient of tropospheric temperature (Kitoh et al., 1997; Sooraj et al., 2015; Turner & Annamalai, 2012; Wang et al., 2021), whereas the ISM is expected to strengthen in terms of rainfall over the country due to enhanced moisture convergence. The warm pool's westward expansion over the AS is consistent with possible weakening of winds over the region.

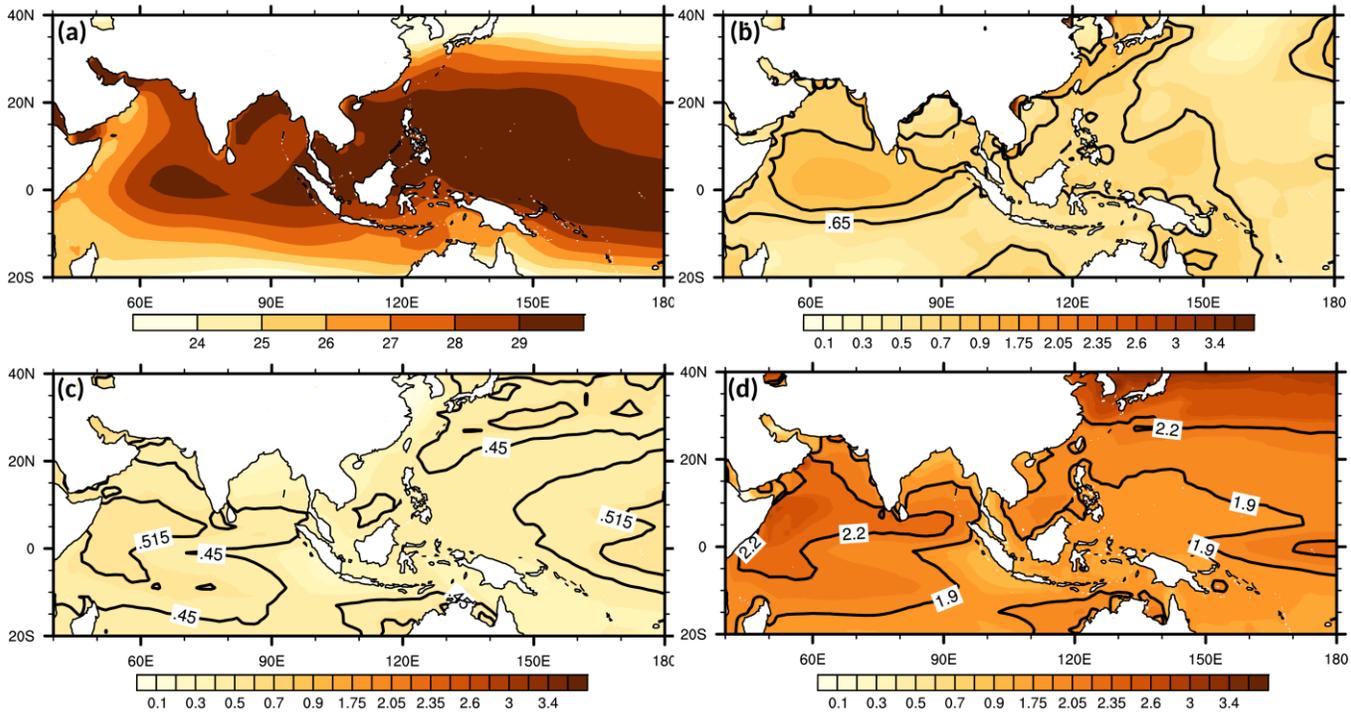

**Fig 2. Observed and simulated changes in the Indian Ocean Warm Pool** (a) Present-day JJAS SST climatology from HadISST (1985–2015) (shading); (b) Observed climatological changes in historical JJAS SST between the present 30-years ($SST_{pres}$) (1985–2015) and the pre-industrial 30–years ($SST_{past}$) (1901–1930); (c) corresponding changes in the simulated ensemble mean SST climatology by 34 CMIP6 historical models between $SST_{pres}$ and $SST_{past}$. (d) The SST changes during the projected period are calculated as the difference between multi-model ensemble mean projected climatology for SSP2–4.5 from the projected period ($SST_{proj}$) (2071–2099) and $SST_{past}$. The thick black SST contour in (c&d) is provided to distinguish the regions between strong warming and relatively less warming. The SNR is greater than one for the entire domain.

4.3 The lengthening rainy season over Northwest India

Using the long-term rainfall observations across the subcontinent(Rajeevan et al., 2010), we verified the westward expansion of the mean ISM rain belt simulated by the CMIP6 historical models using daily rainfall data from the India Meteorology Department (IMD). The change in climatological mean JJAS rainfall between $PR_{past}$ and $PR_{pres}$ indicates a 10–20% increase in mean rainfall over the northwestern part of India; however, a similar decrease in percentage mean rainfall over the eastern part of India (Fig. 3a). The spatial distribution and magnitude of change in land rainfall amounts (mm) are comparable to that of the simulated CMIP6 MME changes (Fig. 1b), with an increase in rainfall in the east and a decrease over the west.

The ISM is a manifestation of the seasonal migration of the ITCZ (Gadgil, 2003; Goswami & Chakravorty, 2017; Webster et al., 1998), with the ISM rainy season marked by a sharp monsoon "onset" over the southern tip of the peninsula, Kerala (MoK), and a comparatively gradual "withdrawal" from Kerala. Even though solar heating produces a heat-low over land by April and May and sets up near-surface cross-equatorial flow, bringing warm, moist air to the continent and accumulating low-level Moist static energy (MSE), the subsidence aloft inhibits convection and prevents the ITCZ from moving beyond 10°N (Xavier et al., 2007). The MoK between the last week of May and the first week of June marks the day when the conditions become conducive for moist symmetric instability to break the inhibition barrier, leading to vigorous ITCZ scale organised convection and the onset of the monsoon over Kerala(B. N. Goswami & Xavier, 2005). Feedback between the latent heating and circulation initiates the first northward propagating pulse of monsoon intraseasonal oscillations (MISO) (Jiang et al., 2004) and sets up monsoon 'onset' at northward locations and monsoon over the whole country by the last week of June or the first week of July. The rainfall during the monsoon season is associated with the ITCZ extending over the whole continent. The east-west asymmetry in the change of seasonal mean rainfall in observations as well as in model simulations during historical period and projections could be due to an asymmetric change in the LRS between the east and west. Based on $1° \times 1°$ daily rainfall data from IMD (Rajeevan et al., 2010) between 1901 and 2019, we calculate the climatological length of the rainy season (LRS) between the monsoon's local "onset", and "withdrawal" dates over the country using a method proposed by Mishra (Misra et al., 2018). The results demonstrate an east-west asymmetry in the mean LRS, ranging from about 150–160 days in north-eastern India to 50–70 days over north-western India (Fig. 3b). As the climatological LRS over northwest India is usually characterised by a comparitively "delayed onset" and "early withdrawal" (Misra et al., 2018) of monsoon, an increase in the LRS can significantly enhance the mean JJAS rainfall over this region. The estimated changes in LRS within the domain between 1901 and 2019 show an increase in the LRS over the NW (Fig. 3c), which is consistent with the changes shown in Fig. 3a, and the spatial distribution and magnitude of change in land rainfall amounts are comparable to the simulated CMIP6 MME changes (Fig. 1b), with an increase in rainfall in the east and a decrease in rainfall in the west.

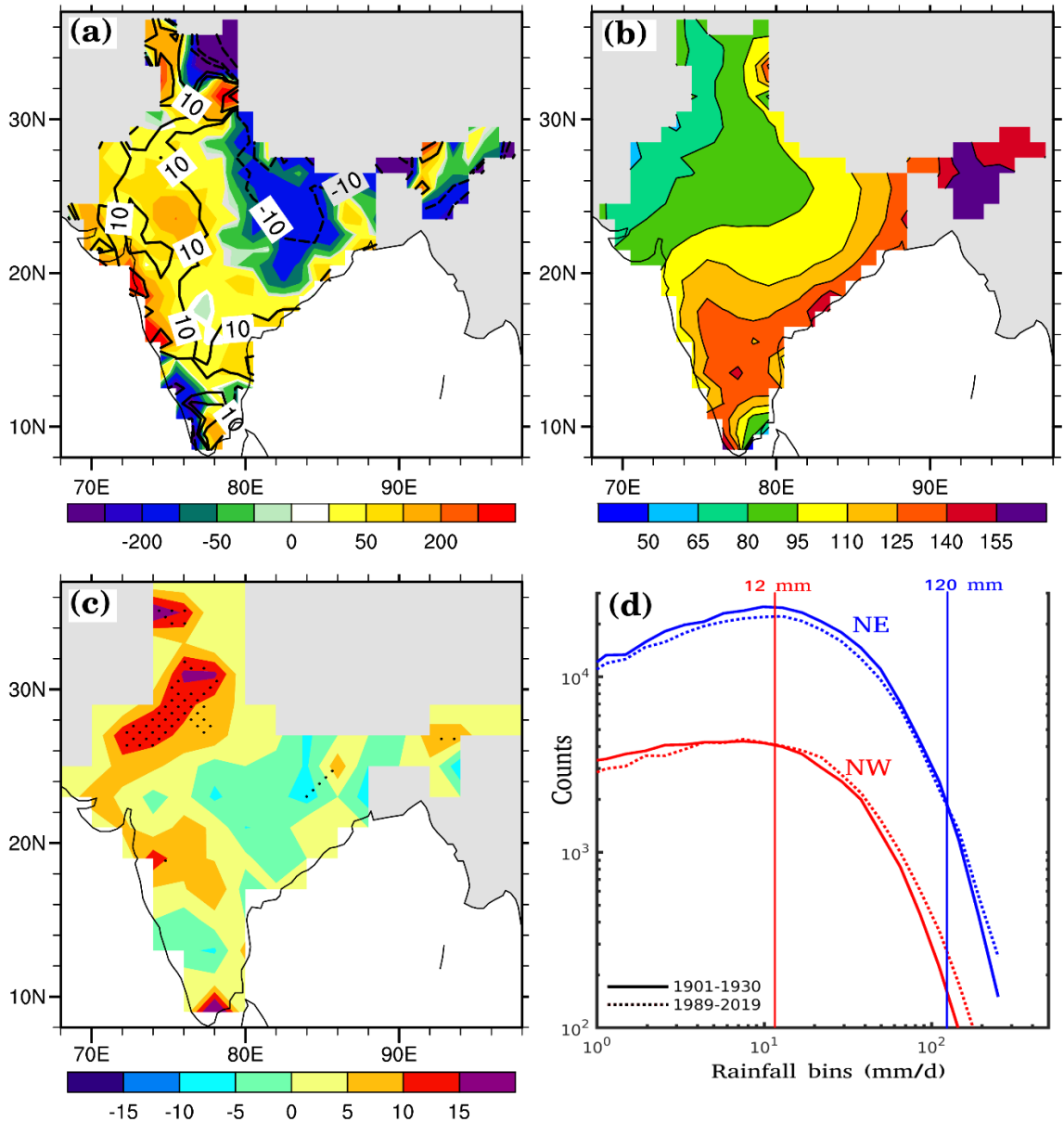

**Fig 3. Changing monsoon rainfall characteristics over the Indian subcontinent** (a) Change in JJAS mean ISM rainfall climatology (mm) over the Indian continent between 1989–2019 and 1901–1930 using IMD daily 1°x1° gridded rainfall data.(b) Long-term climatology of the length of the rainy season (LRS, days) using the Misra et al., 2018 method based on the IMD data set.(c) Change in LRS (days) during the period 1901–2019 based on the trends of LRS, where the dotted regions are statistically significant at 95% CI (d) Daily rainfall probability distributions over central-eastern India (CI, 80E°–95°E, 15N°–25°N) (blue) and northwest India (NWI, 65E°–80°E, 15°N–35°N) (red); from 1901 to 1930 (thick curve) and 1989 to 2019 (dotted curve).

What changes in the precipitating systems led to this increase in seasonal rainfall over northwest India in recent decades? The westward expansion of the IO warm pool is associated with a significant increase in the low-level moisture content of the atmosphere over the AS due to an increase in evaporation as indicated by the increase in the area under the $27°C$ isotherm (Fig. S3a,b). Since the beginning of the century, there has been an increase in integrated moisture transport to northwest India due to the higher moisture loading of the lower atmosphere (Fig. S3c), which is consistent with the increasing trend of cloud base MSE available over northwest India during the recent period 1979 – 2019 (Fig. S3c).

Increased buoyancy and instability over the region are likely to be responsible for the region's increased rainfall during the JJAS season, as well as for an increase in the frequency of mesoscale and synoptic storms which originate there. This is evidenced by an increased number of low-pressure systems (LPS) reaching the region (Fig. S4a), leading to an increase in the number of LPS days over the region (Fig. S4b) and an increased contribution of LPS to total rainfall (Fig. S4c).

Under GW, the frequency of low and moderate rain events (5–100 mm day$^{-1}$) that account for more than 80% of the mean rainfall decreases over central India (CI), also known as the core monsoon domain, while heavy rain events (> 100 mm/day) increases ( Goswami, Madhusoodanan, et al., 2006). The frequency distributions of daily rainfall between 1901–1930 and 1985–2019 over central northeast India (CNE, 80°E–95°E, 15°N–30°N) also show a decrease in low and moderate rain event frequency and an increase in heavy events during recent years, like that found over CI (Goswami, Venugopal, et al., 2006) (Fig. 3d). In contrast, in semi-arid northwest India, the frequency of occurrence of both heavy rain events and low and moderate rain events has increased over the past few decades in comparison to earlier periods (Fig. 3d). This demonstrates that the increase in the mean rainfall over the region (Fig. 3a) results from a qualitative change in the rainfall distribution facilitated by an increase in the frequency of mesoscale and synoptic and intraseasonal disturbances (B. Li et al., 2022) in the region.

## 4.4 South Asian Monsoon Change in the context of Global Monsoon Change

The South Asian monsoon is an integral part of the northern summer ITCZ, or global monsoon. Is the westward expansion of the ISM being a result of expected changes in the global ITCZ under GW? An examination of the simulated ensemble mean climatology of JJAS rainfall (Fig. 4a) and change in JJAS rainfall (Fig. 4b, c, d) over the global tropics between the early historical period and the present period reveals differences in the impact of climate change over the South Asian monsoon region and the South American monsoon system(de Carvalho & Cavalcanti, 2016). In the equatorial Pacific, instead of a westward expansion, there is an eastward expansion of the warm pool and the associated rain belt due to the intensification of the ITCZ with a slight southward shift from its climatological position (Fig. 4 b,c&d). These changes in the rain belt are consistent with the simulated SST changes (Fig. S5), including the increase in the north-south gradient of SST in the eastern Pacific, facilitating the intensification of the ITCZ. The decrease in rainfall over the tropical Americas is associated with the subsidence of the Walker circulation as a Matsuno-Gill response to intense heating over the central and eastern equatorial Pacific to the west. A similar examination of changes in December–January–February (DJF) rainfall climatology (Fig. S6) and SST climatology (Fig. S7) also indicate similar changes in the ITCZ with an eastward expansion and intensification of the equatorial rain belt and a decrease in rainfall over tropical South America. The decrease in rainfall over the tropical South American landmass is between 10–15%, like that over the eastern part of the South Asian monsoon (Fig. 1). However, unlike the south Asian monsoon region, the South American monsoon system is weakened by the eastward expansion of the rain belt over the equatorial Pacific, with no potential benefit from climate change. The westward expansion of the south Asian summer monsoon bringing in transformative changes over northwest India's semi-arid landmass by anthropogenic climate change is, therefore, unique over the global tropics.

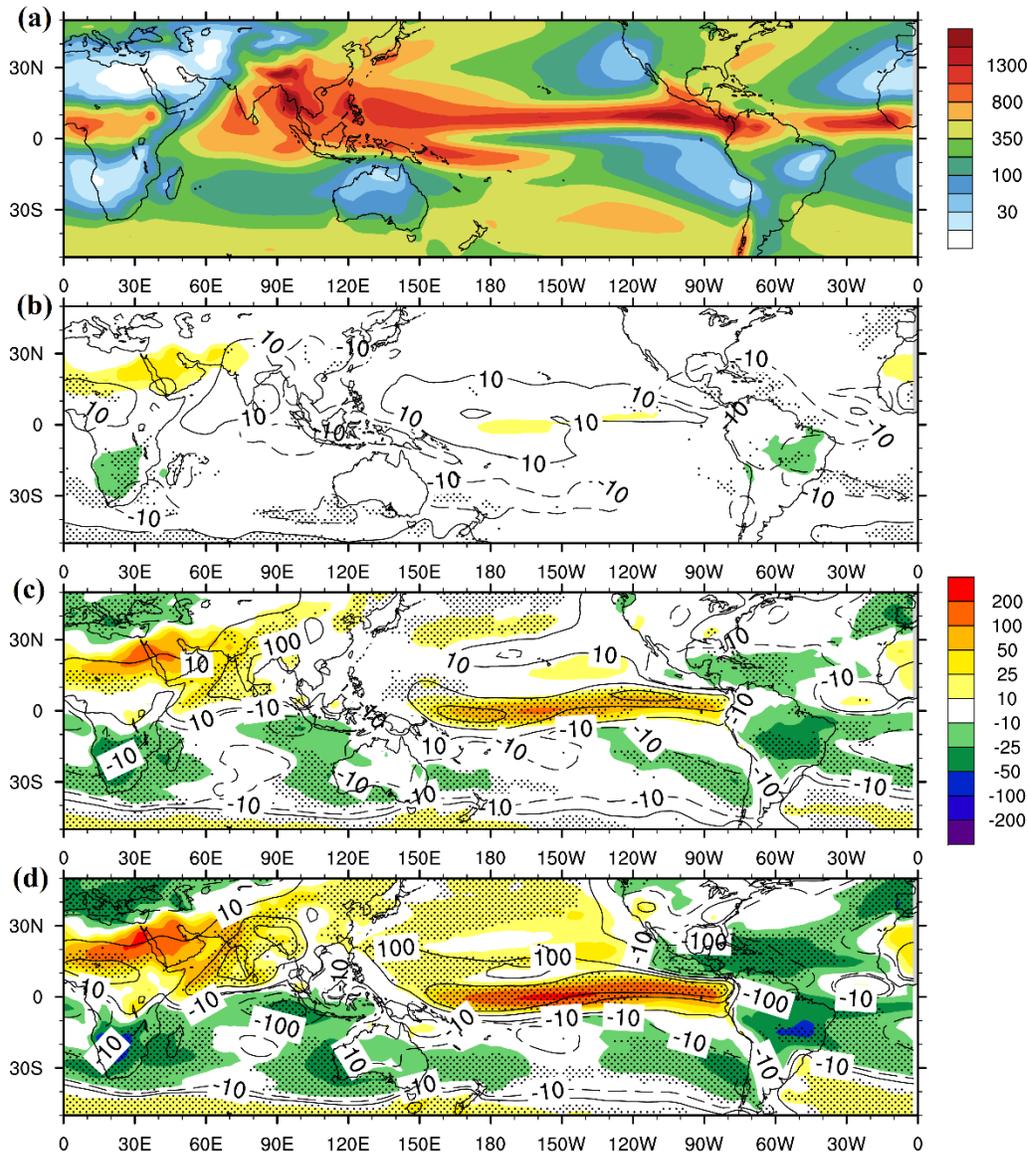

**Fig 4. Changes in simulated boreal summer monsoon rainfall** (a) Global distributions of present-day JJAS MME accumulated rainfall climatology (mm) from CMIP6 historical simulations (shading) (1979–2015), similar to Fig. 1, but shown for the entire tropical region (b) Contours represent rainfall climatology differences calculated as $PR_{pres} - PR_{past}$ during the simulated historical period; (c) rainfall climatology differences calculated as $PR_{proj} - PR_{past}$ for the SSP2–4.5 ensemble; and (d) for the SSP5–8.5 ensemble, respectively. The color shading in (b, c, and d) represents the percentage change (%) in rainfall amount with reference to the past climatology ($PR_{hist}$) and the contours represent the changes in seasonal accumulated rainfall amount (mm).The dotted area denotes the regions where the magnitude of the projected trend (MME mean change) to signal-to-noise ratio (SNR) (inter-model standard deviation) is greater than one.

## 5 Discussion and Conclusions

While there is consensus that the ISM rainfall (Katzenberger et al., 2021; Rajesh et al., 2021; Wang et al., 2021), and the frequency and intensity of the daily extreme rainfall events are expected to increase with the global warming( Goswami, Wu, et al., 2006; Rajeevan et al., n.d.; Rajesh et al., 2021) a major change in the spatial distribution of mean rainfall over the region has been either missed or ignored by studies so far. In contrast to the much hyped 'wet gets wetter and dry gets drier' paradigm, here, using observations and climate model simulations, we show that the mean rainfall over the semi-arid northwest parts of India and Pakistan is going to increase by 50–200% under moderate greenhouse gas increase scenarios (e.g., SSP2–4.5). The westward expansion of the Indian monsoon rainfall is a result of a westward expansion of the summer ITCZ, facilitated by a westward expansion of the IO warm pool. The wetting of northwest India primarily happens in the northern summer, with no signal of increasing rainfall over the region during the northern winter (DJF) due to extratropical influence (Fig. S10) and westward expansion of the ITCZ only benefiting east-central Africa. Equilibrium coupled climate model simulations with 4xCO2 (Fig. S8, Table S2) compared to responses to pre-industrial CO2 concentrations indicate that the westward expansion of the Indian monsoon rainfall is basically driven by the greenhouse gas (GHG) forcing. The observed and projected westward expansion of the summer ITCZ appears to be caused by ocean-atmosphere coupled feedback, in which atmospheric circulation changes caused by GHG forcing drive a westward expansion of the warm pool in the IO, favoring intensification of the westward expansion of the ITCZ, and the northward excursion of the ITCZ in this region results in the westward expansion of the ISM.

There is a large east-west asymmetry in the sensitivity of seasonal rainfall across the Asian monsoon region. The change in climatological JJAS mean rainfall between past thirty years (PR$_{past}$) and present 30 years (PR$_{pres}$) from the historical period (1986–2015), normalized by global mean surface temperature change (Fig.S9a) indicates that the rainfall over the northwestern part is increasing at 100%/K, the central part is increasing at 9%/K while the eastern part if decreasing at 8%/K. Under SSP4.6 (SSP8.5) the change between the past 30 years (PR$_{past}$) and last 30 years of the century (2071–2100) from projections ($PR_{proj}$) normalized by global mean temperature change during the period (Fig. S9b,c) shows a 70%/K increase in the northwest, a 8%/K increase in the central part and a 1.5%/K increase in the eastern part. With general consensus on a 2–3 degree increase in global mean temperature, a 2.5 fold increase in the climatological rainfall during JJAS is a good estimate of rainfall increase by the end of the century in the northwestern part of the South Asian Monsoon. Although the climatological mean rainfall simulation over the ISM region has steadily improved from the CMIP3 to CMIP6 generations of climate models, a large fraction of CMIP6 models still have significant dry biases in simulating the JJAS rainfall over land as well as SST over the IO (Choudhury et al., 2021). Therefore, individual models' biases may impact the signal of the westward extension of monsoon rainfall and that of the IO warm pool. A moderately strong sign of the westward expansion of rainfall as well as that of a warm pool in the multi-model ensemble mean climatology indicates the robustness of the signal. To determine how these biases could affect the strength of the signal, we compared the westward expansion of JJAS rainfall and JJAS warm pool SST in the multi-model ensemble mean CMIP5 models (Figs. S10 and S11) with that of CMIP6 models (Fig. 1 and Fig. 2). We notice that the spatial pattern of rainfall change simulated by CMIP5 models is somewhat similar to that simulated by CMIP6 models, but the amplitude of the increase in rainfall over northwest India is only about half that by CMIP6 models, with poor inter-model SNR over the NW. However, CMIP5 models closely resemble CMIP6 models in terms of the overall pattern and amount of the SST changes. The weaker changes in rainfall in CMIP5 models over northwest India are therefore likely caused by the cold bias of the climatological mean SST compared to CMIP6 models, even with a comparable increase in SST in both sets of models.

As the ensemble mean JJAS rainfall climatology simulated by the CMIP6 models has a dry bias over central and western India (Fig. S1b), the intensity of the westward expansion of monsoon rainfall from the CMIP6 models (Fig. 1a) could still be an underestimate. Even the "best" 10 CMIP6 models (Choudhury et al., 2021) that adequately simulate the climatological mean JJAS rainfall monsoon exhibit consistent bias towards the south when simulating the land ITCZ. Our analysis using the "best" 10 models with significantly strong spatial correlations between simulated and observed ISM climatology (Fig. S12) could also demonstrates a similar increase in rainfall over northwest India in compared to the multi-model ensemble mean of 34 CMIP6 models (Fig. 1b). Although the 34 models employed in this study may still have varied degrees of biases in simulating the JJAS mean rainfall climatology over the ISM region(Choudhury et al., 2021), the CMIP6 models have improved greatly compared to their predecessors (CMIP3, CMIP5).

With significant inter-model variations, how reliable is the ensemble mean of the 34 models and how reliable is the change of the simulated ensemble mean? The inter-model spread is largely due to model's systematic biases in simulating the ISM change, as well as their biases in simulating global temperature change (Rajesh et al., 2021). Several studies have now demonstrated that the inter-model spread could be reduced and confidence in multi-model mean simulations of ISM could be enhanced by correcting ISM simulations using emergent constraints (EC) (G. Li et al., 2017; Shamal & Sanjay, 2021). Recently, we have unraveled an EC (Rajesh & Goswami, 2022) for correcting the ISM projections and shown that the inter-model spread could be reduced by 14%. The fact that even after applying the EC, the westward extension of monsoon rainfall remains similar or is even slightly enhanced with the same SNR provides evidence of the signal's robustness (Fig. S13). We see that amplified rainfall changes across deep Arabia, up to 200–300% change, are unreliable because of the low SNR in the region, despite being significant relative to their base level.

With no clear commitments to drastically cut down GHG emissions, most studies indicate that the global mean temperature will increase by more than 2°C by the end of the century (Rogelj et al., 2016; Sognnaes et al., 2021). In this backdrop, our study indicates a high probability of an increase of 2.5 fold in mean rainfall during the JJAS season over northwest India by the end of the century. In any semi-arid climate, total seasonal rainfall comes from a limited number of short and intense rain events that lead to increased frequency of hydrological disasters. Our finding that the frequency of extreme rainfall events is increasing in the northwest India in recent years (Fig.3d) is supported by recent study (Sreenath et al., 2022) that finds that such events are increasing from deeper convective activity in the west coast of India in recent years. The great flood event of Pakistan of 2022(European Space Agency, 2022)  and those over Gujarat (Davies, 2022), Rajasthan (Economictimes, 2022) and Tamil Nadu (TWC India, 2022)  during summer of 2022 and increased frequency of these hydrological disasters in the central and western parts of India in recent years is a pointer that the process of increased rainfall in the region is underway in earnest. While in the long term, the increased rainfall has the potential of greening the deserts and significantly increasing the food productivity in the region, in the short-term adaptation strategy to the increased intense rain spells and associated hydrological disasters is a must. However, intense spells of rain also tend to run off faster. To effectively harvest the potential of increased rainfall, it is also imperative to plan for harvesting the excess water so as it increases the groundwater reserves in the region

**Acknowledgments**
BNG is grateful to the Science and Engineering Research Board (SERB), Government of India for a Fellowship and a Research Grant. He is also grateful to Debasis Sengupta for valuable suggestions for improvement of the manuscript and Cotton University for hosting the SERB Fellowship. PVR thanks SERB for financial support during the initial phase of this work, and Ministry of Earth Sciences and IITM, Pune for continuing the support.

**Funding:** DST | Science and Engineering Research Board (SERB): Bhupendra Nath Goswami Diary No. SERBIF/ 3707 12020-21, SERB Distinguished Fellowship

**Competing interests:** The authors declare that they have no known competing financial interests or personal relationships that could have appeared to influence the work reported in this paper.

Supporting Information for

# Climate Change and Potential Demise of the Indian Deserts

P. V. Rajesh[1] and B. N. Goswami[2]*.

[1]Indian Institute of Tropical Meteorology, Pune 411008, India
[2]Department of Physics, Cotton University, Guwahati 781001, India

*Corresponding author, email bhupengoswami100@gmail.com

**This PDF file includes:**

Figures. S1 to S13
Tables S1 to S2

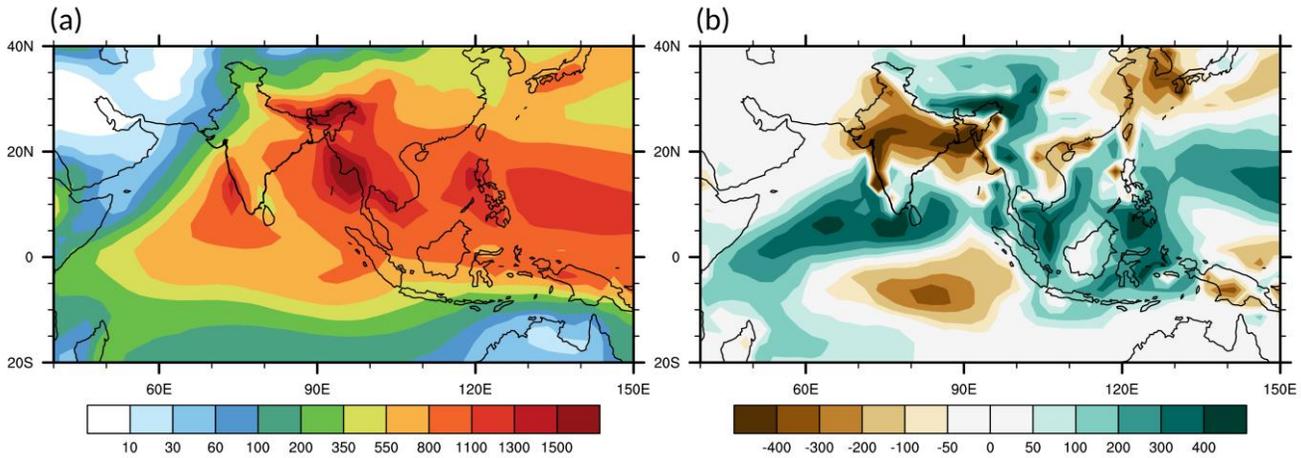

**Fig.S1|** Climatology and bias of summer monsoon rainfall from CMIP6 historical MME (a) Spatial distribution of CMIP6 multi-model ensemble mean JJAS rainfall climatology simulations of the Indian monsoon during the period (1979–2015) by 34 CMIP6 models, similar period to the one as the observed climatology shown in Fig.1a. (b) Bias in simulating the ensemble mean JJAS rainfall climatology by the CMIP6 models.

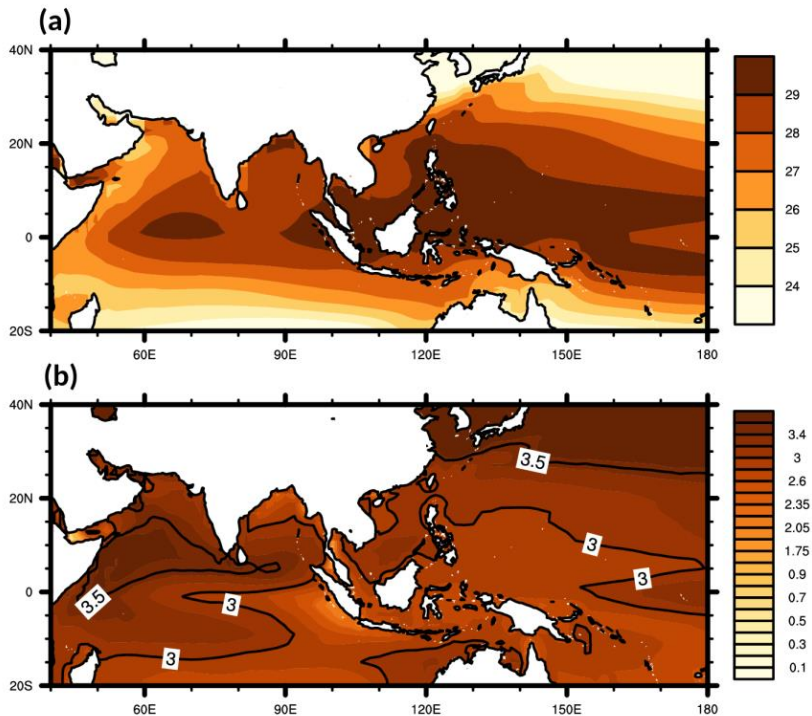

**Fig.S2 |** (a) Mean JJAS SST MME climatology (1979–2015) of CMIP6 historical models and (b) the change in projected JJAS SST MME climatology ( $^{O}$C) by 34 CMIP6 models by the end of 21$^{st}$ century (2071–2099) compared with the past historical climatology (1901–1930) (SST$_{past}$) under SSP8.5.

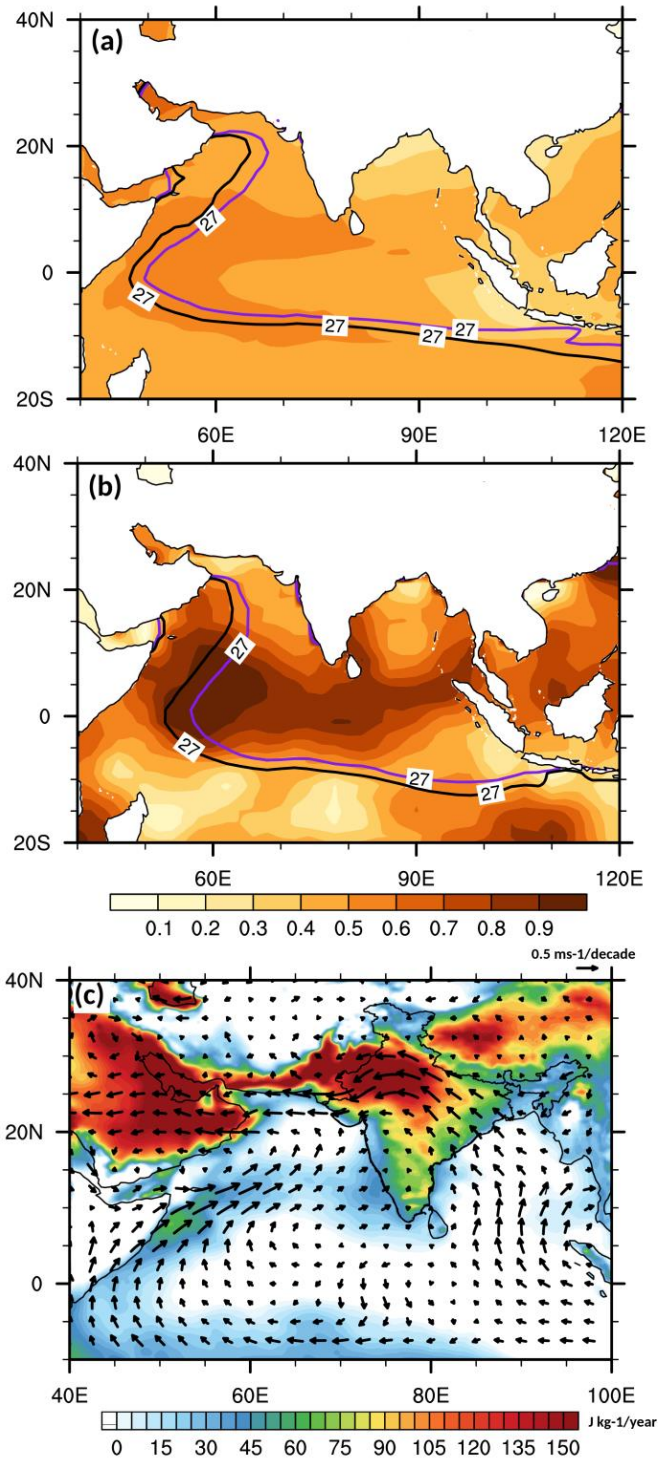

**Fig.S3 |** (a) Simulated changes in JJAS SST MME climatology during historical period by 34 CMIP6 models over the Indian monsoon region (shaded) and the comparison of the changes in spatial extend of the distribution of 27°C isotherm during the past historical period (1901–1930) (violet line) with that of the present period (1985–2015) (black line). (b) Same as (a), but for observed changes for the same period from COBE SST data. (c) Trends of below-cloud-base (950 hPa) moist static energy (MSE) and 850 hPa winds during 1979–2019 obtained from ERA5 reanalysis.

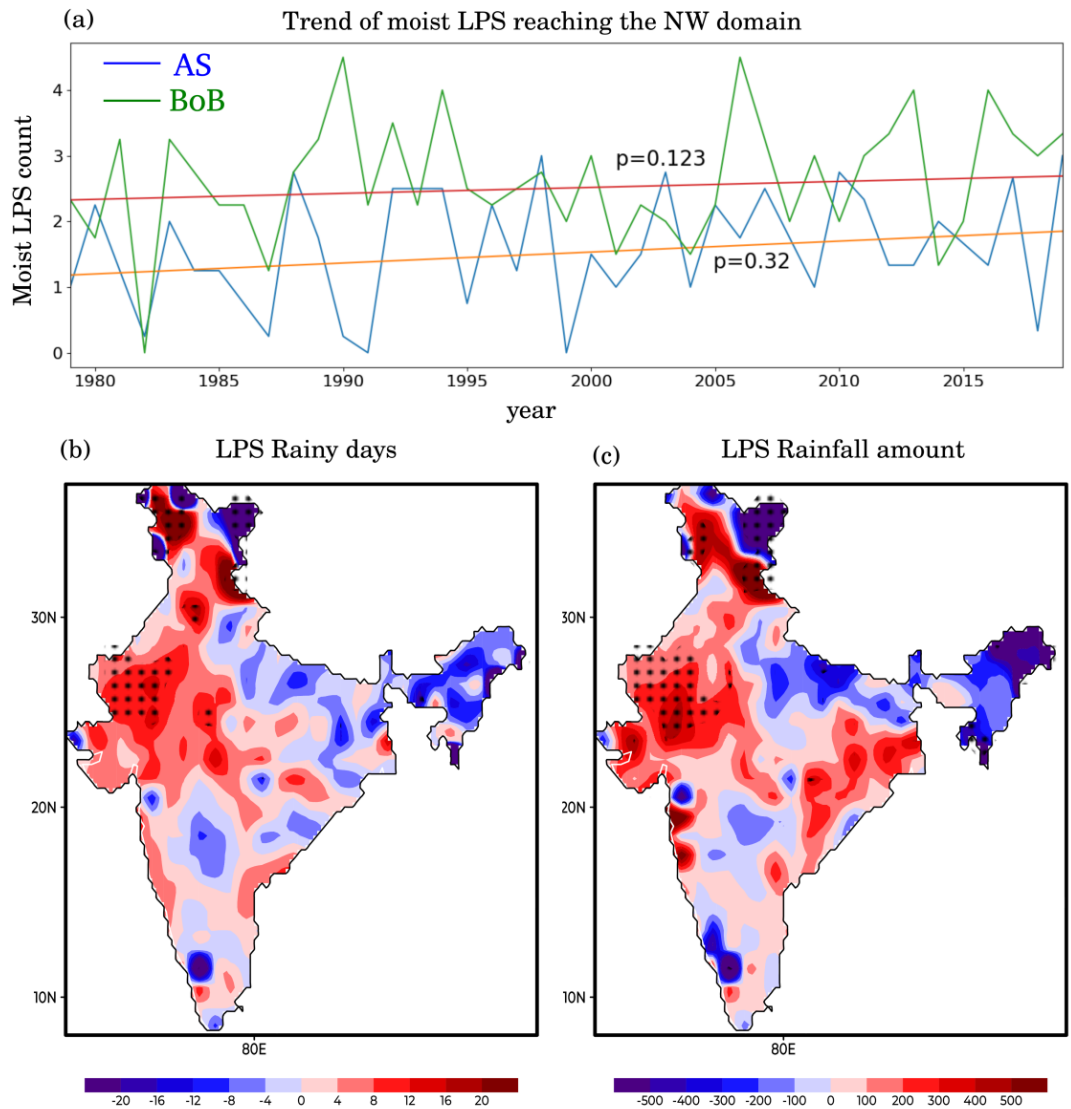

**Fig.S4 |** (a) The changes in the LPS counts, for systems that originate from the Bay of Bengal (BoB) (green) and AS (blue), which reach the deep interior (NW) part of Indian subcontinent (domain 70°E–76°E,20°N–30°N) within the last four decades. The LPS data is obtained from the Global Track dataset of monsoon LPS (Vishnu[4](Vishnu et al., 2020), ERA reanalysis). Using the LPS data, the corresponding rainfall amount and LPS duration are derived using IMD 1 degree data. (b) The trend of total LPS rainy days from 1979 to 2015, and (c) accumulated rainfall amount (mm) from moist low-pressure systems (LPS) originating from both the AS and the BoB during the JJAS season.

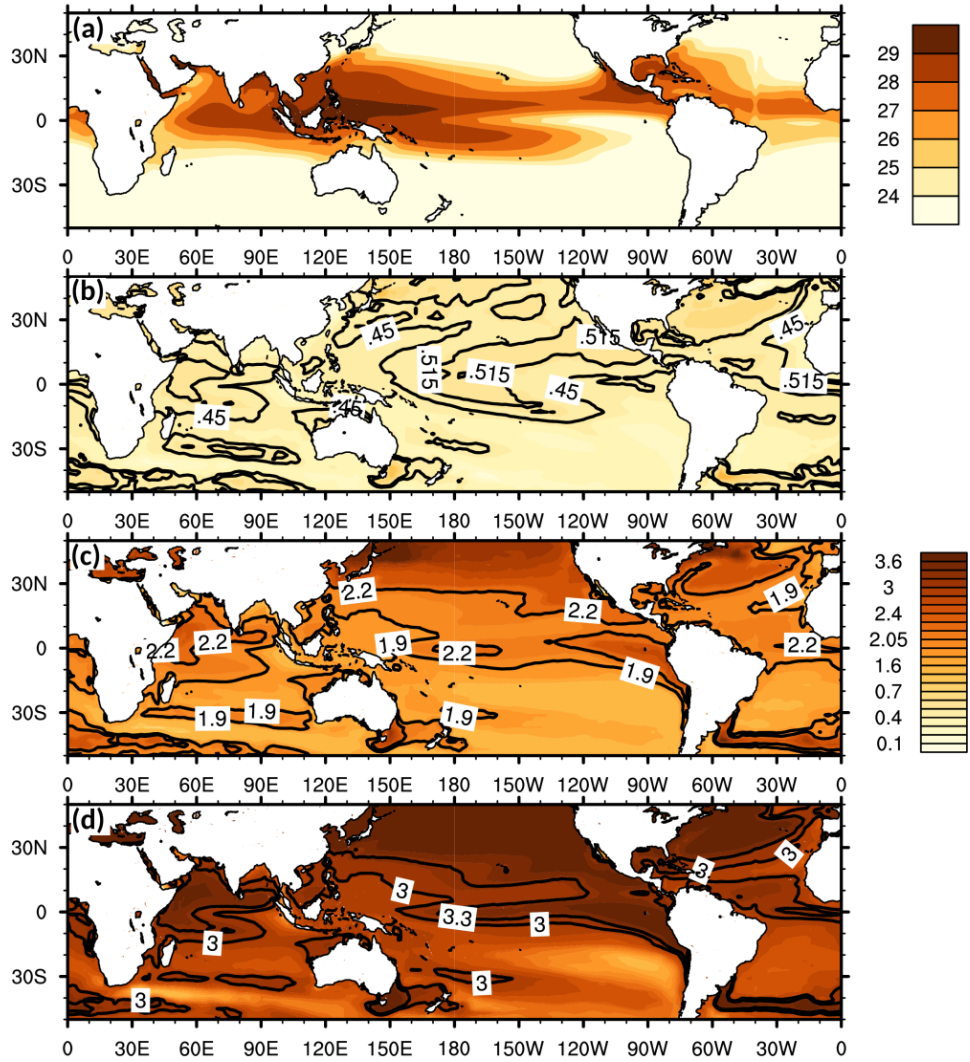

**Fig.S5 |** Same as Fig. 4, but for CMIP6 simulated JJAS SST MME. (a) is the historical climatology, and (b,c, and d) are the changes in SST for historical, SSP2–4.5 and SSP5–8.5, respectively. The contours distinguish the regions between strong warming and relatively less warming. The SNR is >1 for the entire domain.

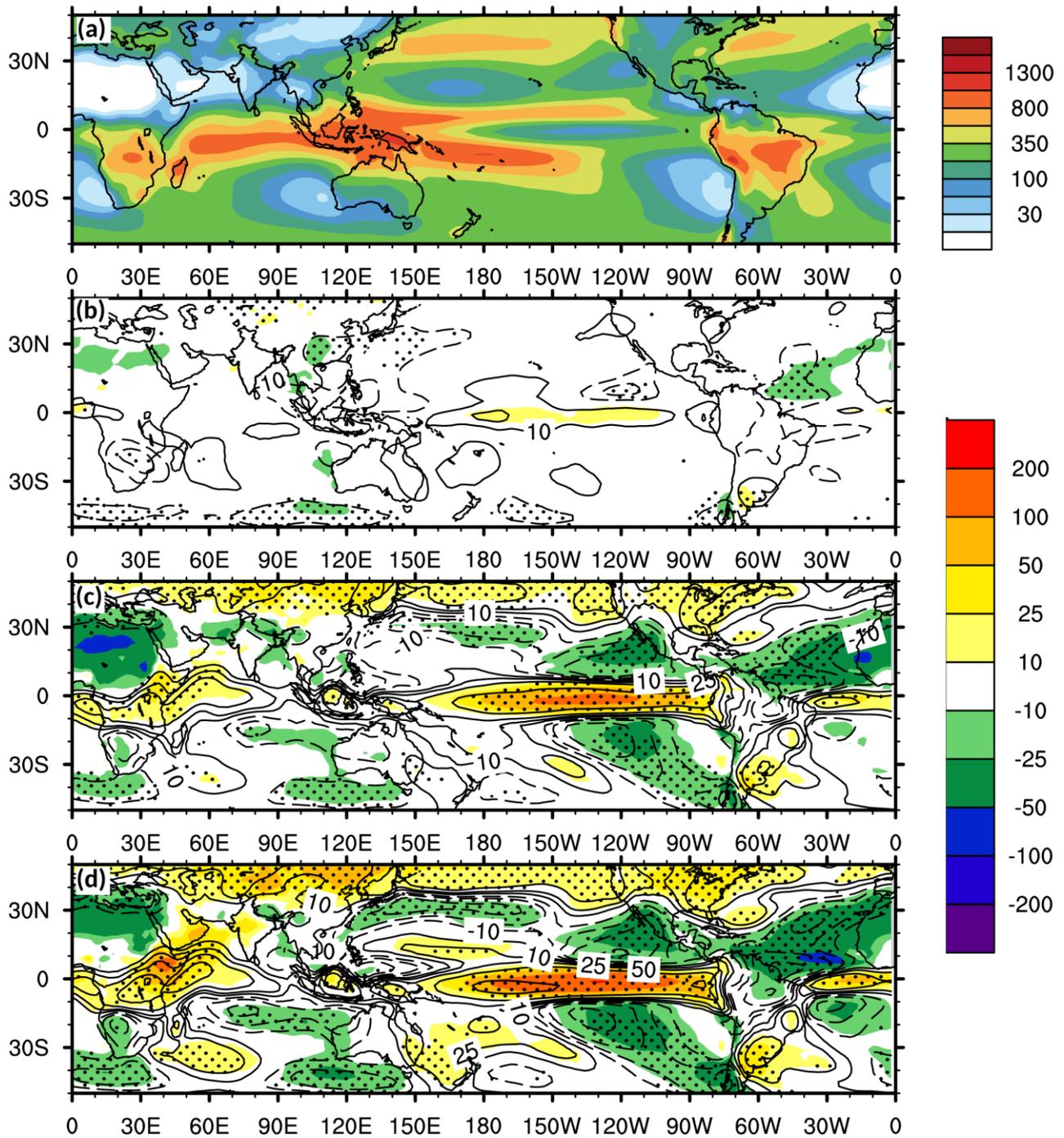

**Fig.S6 |** Same as in Fig. 4, but for DJF season

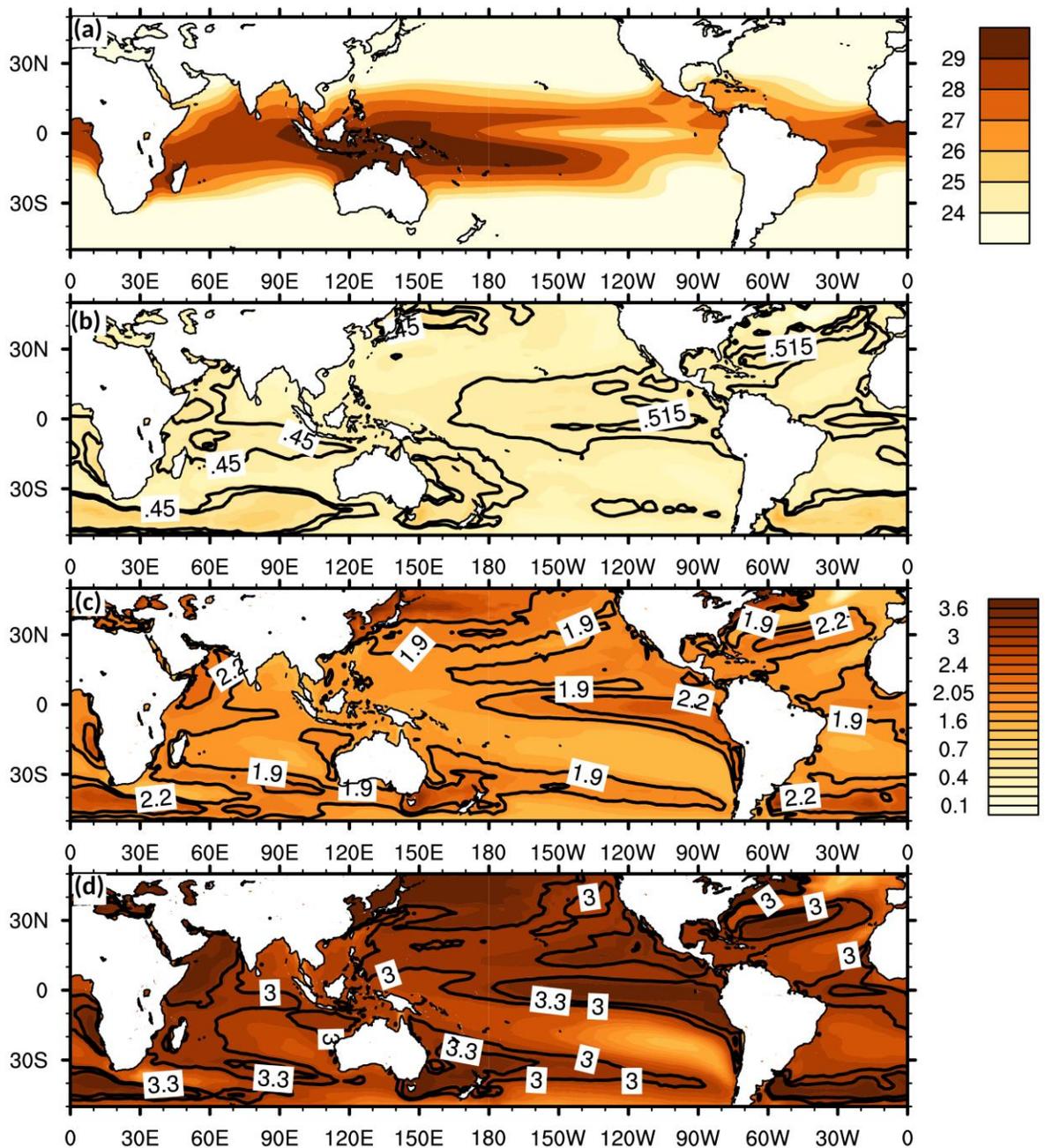

**Fig.S7**|Same as in Fig.S5, but for DJF season

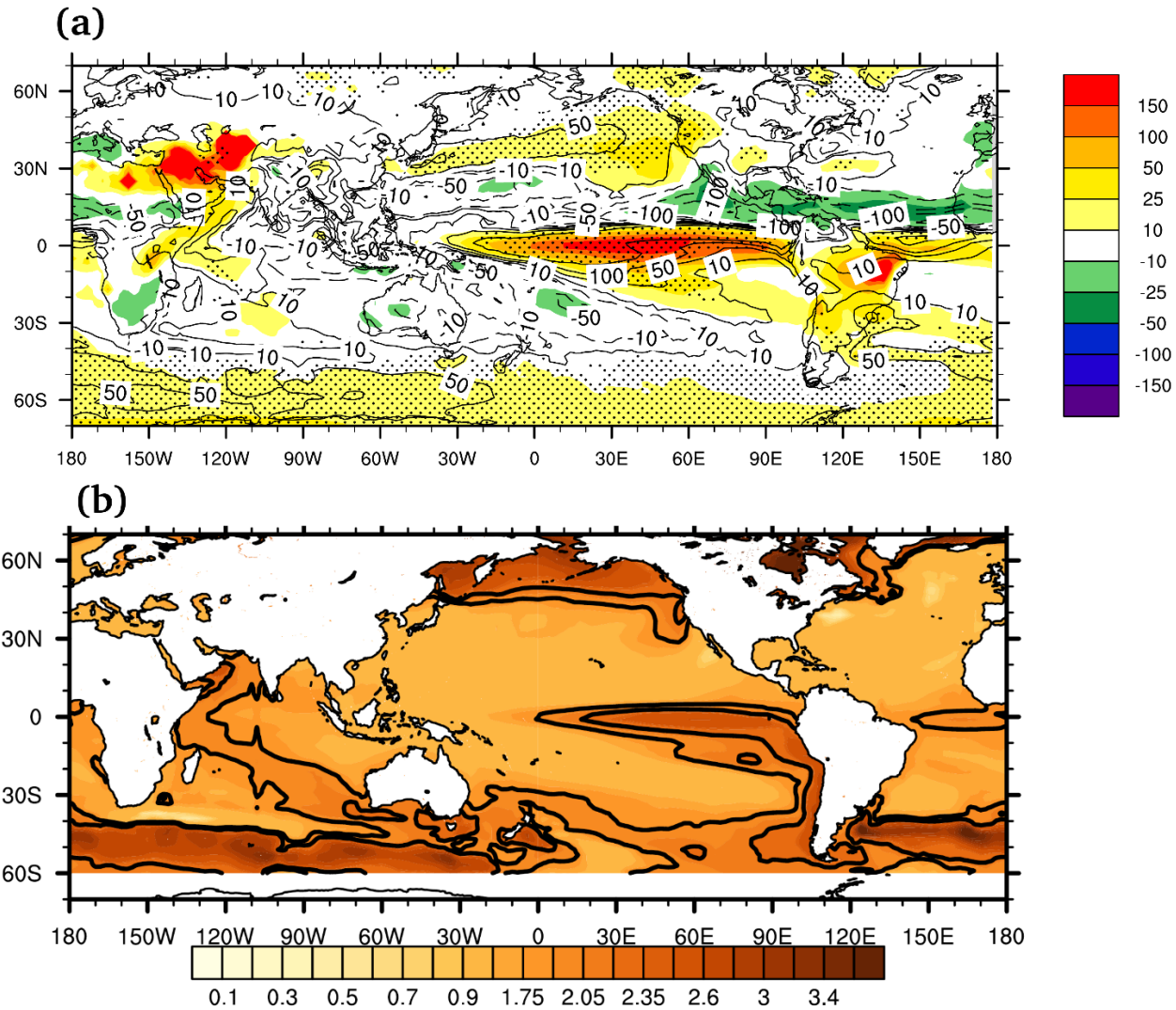

**Fig.S8 | Changes in the (a) rainfall (%) and (b) SST from the ensemble of simulations from abrupt 4xco2 experiments for a 150 year run. Here the % changes in rainfall is calculated as the changes (%) from an initial 30 year climatology with the last 30 year climatology. SST changes (absolute) are also made by this method. The abrupt-4xCO2 scenario uses fully coupled AOGCM model experiments where atmospheric CO2concentrations are rapidly quadrupled from a near-equilibrium, preindustrial state concentration (4 x 285 ppm) and held constant. In this series of equilibrium sensitivity tests, the temperature change was observed after allowing the climate system to stabilise with a greater CO2 level. More pertinent to the changes we are expected to experience in the twenty-first century is the response on shorter time periods, before the deep seas have had time to stabilise. We also intend to minimize the influence of internal variability in evaluating long-term precipitation changes during its evolution by combining thermodynamic/dynamic and cloud radiative feedback processes with the abrupt-4xCO2 scenario.**

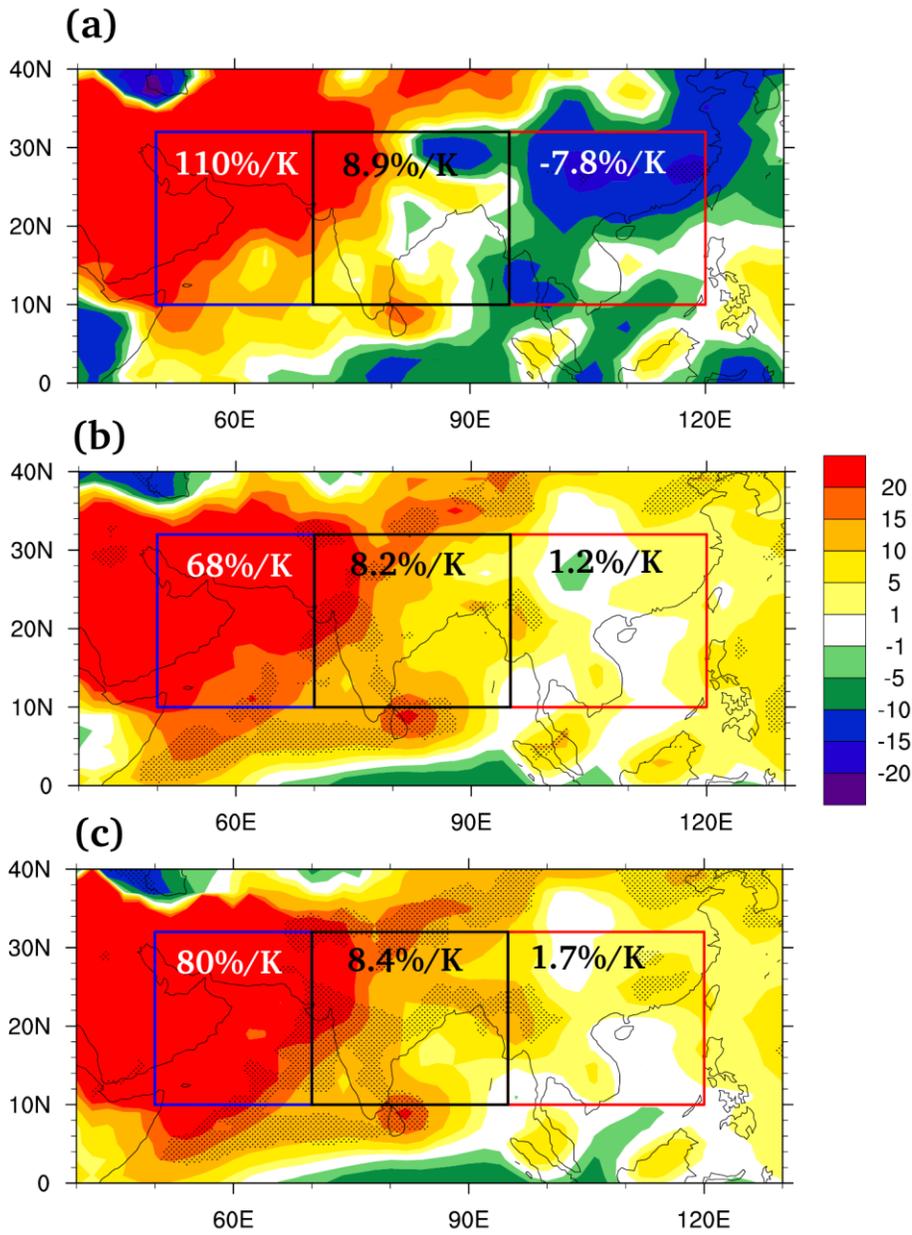

**Fig.S9 |** Hydrological sensitivity (% K$^{-1}$) of (a) historical, (b) SSP2–4.5 and (c) SSP5–8.5 CMIP6 models. Ensemble mean are obtained by normalizing the sensitivity of individual models with their own global mean temperature changes during the respective periods.

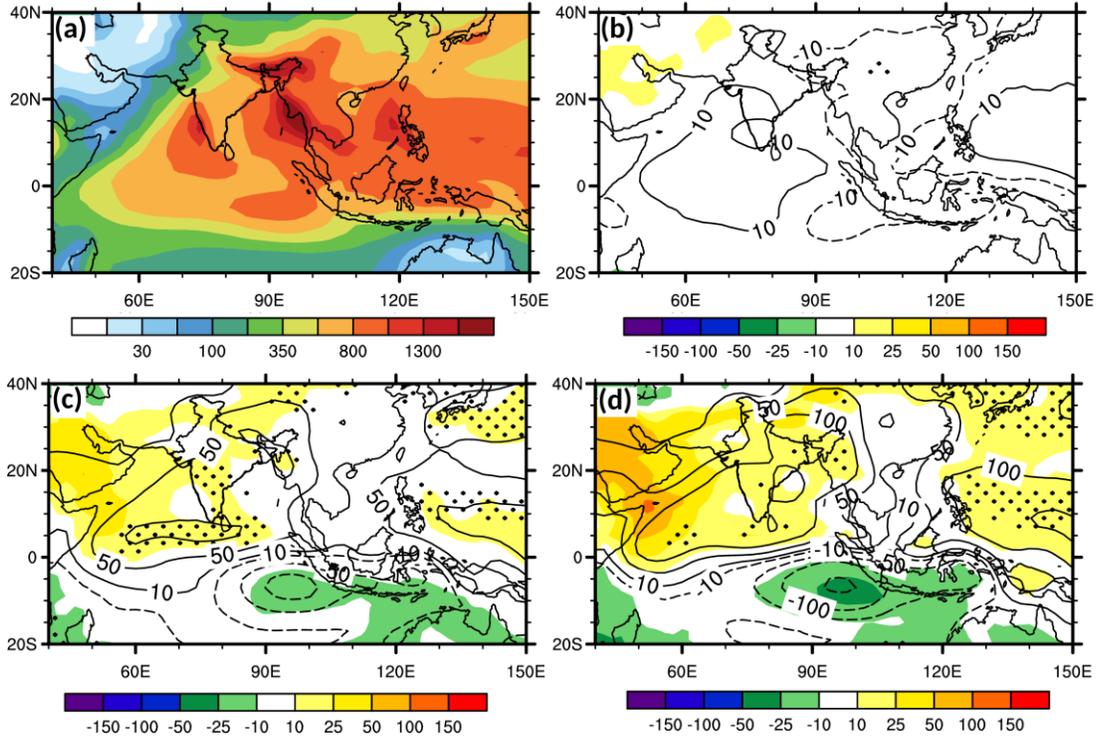

**Fig.S10** | Same as Figure.1 but for 30 CMIP5 MME rainfall

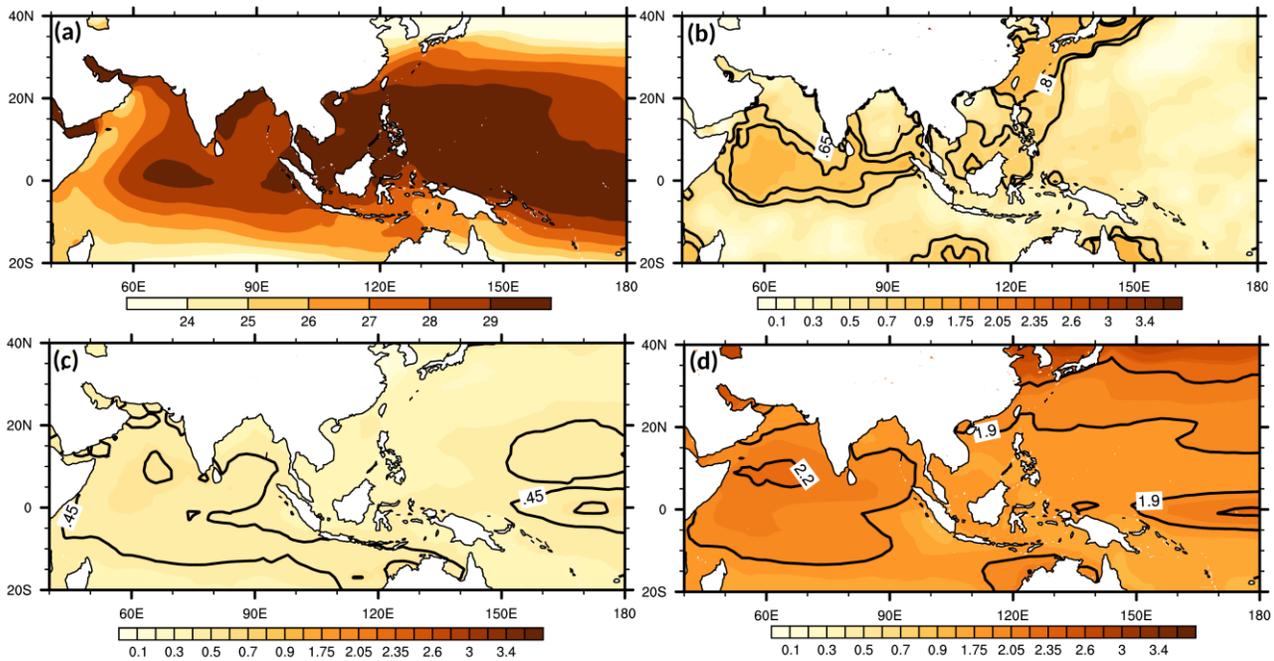

**Fig.S11** |Same as Fig.2 but for CMIP5 MME SST

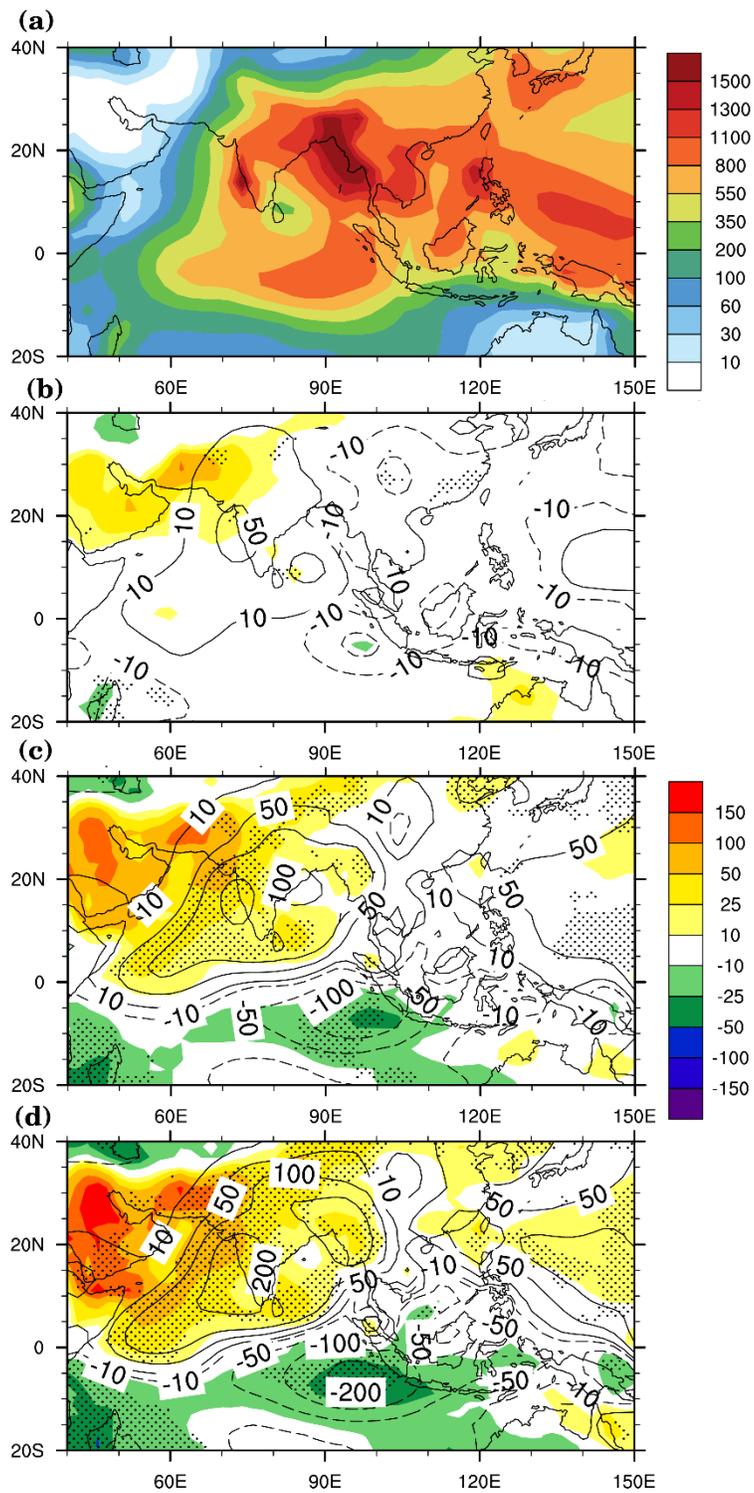

**Fig.S12**| Same as Fig. 1 but for the MME of the "best" 10 CMIP6 models having correlation coefficients greater than 0.75, between the simulated and observed climatological mean rainfall over the Indian monsoon region.

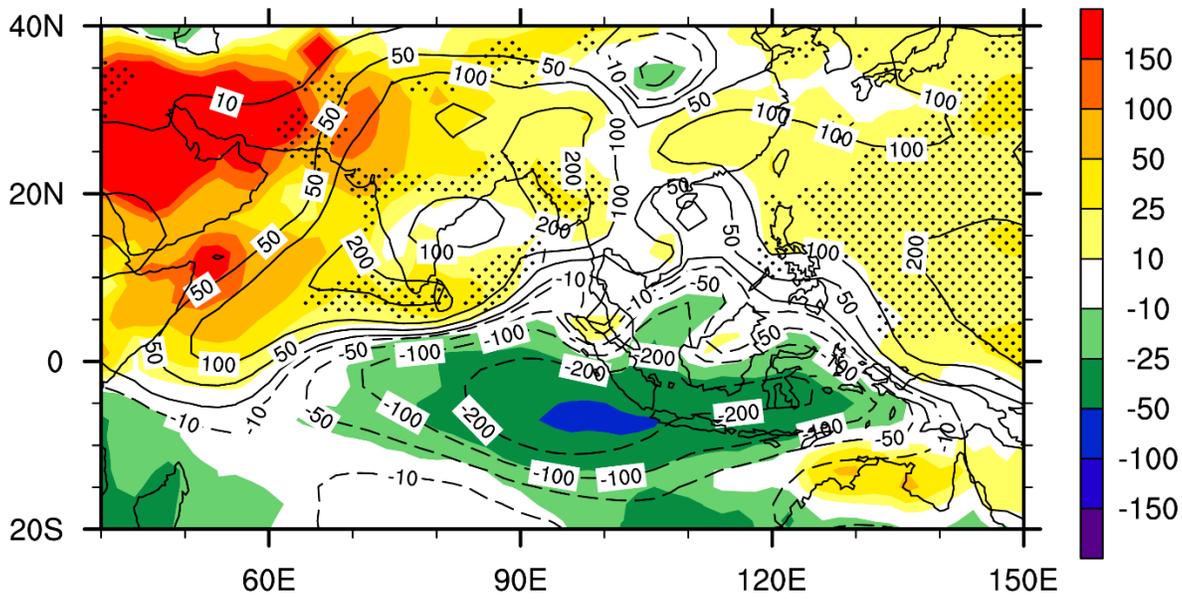

**Fig.S13**| The same as in Fig. 1d, but with corrections applied to individual models (28 models) before computing the MME changes, based on the EC through the relationships discovered between projected changes in the Indian summer monsoon and the multidecadal mode of the Indian summer monsoon (refer to Rajesh and Goswami, 2022 for more details on the EC).

**Table S1 | List of CMIP models used in the study**. Common historical/SSP/RCP models used to construct the MME precipitation/SST fields. A few models from this list are excluded when deriving the MME SST, since the SST fields are not available for those models (not common for all scenarios).

| No | CMIP6 models | No | CMIP5 models |
|---|---|---|---|
| 1 | ACCESS-CM2 | 1 | ACCESS1-0 |
| 2 | ACCESS-ESM1-5 | 2 | BNU-ESM |
| 3 | AWI-CM-1-1-MR | 3 | CCSM4 |
| ↧4 | BCC-CSM2-MR | 4 | CESM1-BGC |
| 5 | CAMS-CSM1-0 | 5 | CESM1-CAM5 |
| 6 | CESM2 | 6 | CNRM-CM5 |
| ♦↧7 | CESM2-WACCM | 7 | CSIRO-Mk3-6-0 |
| 8 | CIESM | 9 | CanESM2 |
| 9 | CMCC-CM2-SR5 | 10 | FGOALS_g2 |
| ↧10 | CMCC-ESM2 | 11 | FIO-ESM |
| ↧11 | CanESM5 | 12 | GFDL-CM3 |
| ♦12 | EC-Earth3 | 13 | GFDL-ESM2G |
| ♦13 | EC-Earth3-Veg | 14 | GFDL-ESM2M |
| ↧14 | FGOALS-f3-L | 15 | GISS-E2-H |
| 15 | FGOALS-g3 | 16 | GISS-E2-R |
| 16 | FIO-ESM-2-0 | 17 | HadGEM2-AO |
| ♦17 | GFDL-CM4 | 18 | HadGEM2-CC |
| ♦18 | GFDL-ESM4 | 19 | HadGEM2-ES |
| 19 | GISS-E2-1-G | 20 | IPSL-CM5A-LR |
| 20 | IITM-ESM | 21 | MIROC5 |
| 21 | INM-CM4-8 | 22 | MIROC-ESM |
| 22 | INM-CM5-0 | 23 | MIROC-ESM-CHEM |
| 23 | IPSL-CM6A-LR | 24 | MPI-ESM-LR |
| ♦24 | KACE-1-0-G | 25 | MPI-ESM-MR |
| ♦↧25 | KIOST-ESM | 26 | MRI-CGCM3 |
| 26 | MCM-UA-1-0 | 27 | NorESM1-M |
| ♦27 | MIROC6 | 28 | NorESM1-ME |
| ♦↧28 | MPI-ESM1-2-HR | 29 | bcc-csm1-1 |
| 29 | MPI-ESM1-2-LR | 30 | bcc-csm1-1-m |
| 30 | MRI-ESM2-0 | | |
| 31 | NESM3 | | |
| ♦32 | NorESM2-LM | | |
| ♦33 | NorESM2-MM | | |
| 34 | TaiESM1 | | |

♦ List of the top 10 models that most closely resemble the observed rainfall climatology (pattern correlation > 0.75)

↧ List of the CMIP6 models that is not included in EC calculations

**Table S2 | List of CMIP models used from abrupt 4xco2 experiments**

| No | CMIP6 models |
|---|---|
| 1 | ACCESS-CM2 |
| 2 | CanESM5 |
| 3 | CESM2 |
| 4 | CESM2-FV2 |
| 5 | CMCC-ESM2 |
| 6 | CNRM-CM6-1 |
| 7 | E3SM-1-0 |
| 8 | EC-Earth3 |
| 9 | FGOALS-f3-L |
| 10 | GISS-E2-1-G |
| 11 | GISS-E2-1-H |
| 12 | IPSL-CM6A-LR |
| 13 | KACE-1-0-G |
| 14 | MCM-UA-1-0 |
| 15 | MIROC6 |
| 16 | MPI-ESM-1-2-HAM |
| 17 | MPI-ESM1-2-LR |
| 18 | MRI-ESM2-0 |
| 19 | NESM3 |
| 20 | NorCPM1 |
| 21 | NorESM2-MM |